\documentclass[preprint,showpacs,superscriptaddress]{revtex4-1}
\usepackage{hyperref}
\usepackage{graphicx}
\usepackage{amsmath,amssymb,amsfonts}

\begin{document}

\title{Physical properties of V$_{1-x}$Ti$_{x}$O$_{2}$ (0 $<$ x $<$ 0.187) single crystals}

\author{Tai Kong}
\affiliation{Ames Laboratory, US DOE, Iowa State University, Ames, Iowa 50011, USA}
\affiliation{Department of Physics and Astronomy, Iowa State University, Ames, Iowa 50011, USA}
\author{Morgan W. Masters}
\affiliation{Department of Physics and Astronomy, Iowa State University, Ames, Iowa 50011, USA}
\author{Sergey L. Bud'ko}
\affiliation{Ames Laboratory, US DOE, Iowa State University, Ames, Iowa 50011, USA}
\affiliation{Department of Physics and Astronomy, Iowa State University, Ames, Iowa 50011, USA}
\author{Paul C. Canfield}
\affiliation{Ames Laboratory, US DOE, Iowa State University, Ames, Iowa 50011, USA}
\affiliation{Department of Physics and Astronomy, Iowa State University, Ames, Iowa 50011, USA}

\begin{abstract}
Free standing, low strain, single crystals of pure and titanium doped VO$_{2}$ were grown out of an excess of V$_{2}$O$_{5}$ using high temperature solution growth techniques. At $T_{MI} \sim$ 340 K, pure VO$_{2}$ exhibits a clear first-order phase transition from a high-temperature paramagnetic tetragonal phase (R) to a low-temperature non-magnetic monoclinic phase (M1). With Ti doping, another monoclinic phase (M2) emerges between the R and M1 phases. The phase transition temperature between R and M2 increases with increasing Ti doping while the transition temperature between M2 and M1 decreases. 
\end{abstract}
\maketitle

\begin{figure}[!h]

\includegraphics[width=0.9\textwidth]{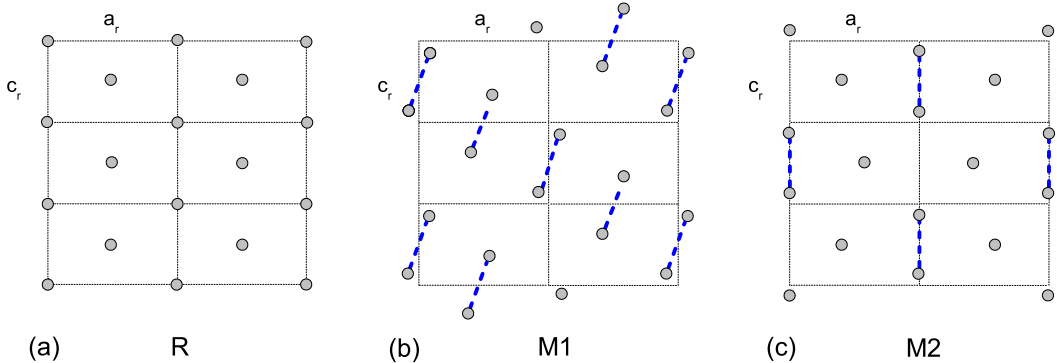}

\caption{(Color online) (a), (b) and (c) show schematics of V-V pairing in R, M1 and M2 phases respectively (The distortions are exaggerated for clarity). Solid circles represent V$^{4+}$ ions. In (b) and (c), the V$^{4+}$ ions connected by blue dashed lines are dimerized. From Ref.~\onlinecite{Marezio72}.}
\label{structure}
\end{figure}

The metal-insulator (MI) transition in VO$_{2}$ at around 340 K was first reported by Morin in late 1950s\cite{Morin59}. Ever since, great effort has been made to understand the mechanism behind this MI transition as well as to explore its potential application in electronic devices\cite{Imada98,Yang11}. Samples in various forms have been synthesized: bulk (polycrystalline and single-crystalline)\cite{Morin59,Bongers65,Sasaki64,MacChesney69,Ladd69}, thin films and nano-structures\cite{Yang11}. At high-temperatures, VO$_{2}$ is in a paramagnetic state with a tetragonal (P4$_{2}$/mnm) rutile structure (R). Below the MI transition the V$^{4+}$ ions dimerize into non-magnetic pairs and cant/twist into a monoclinic (P2$_{1}$/c) structure (M1)\cite{Imada98}. Fig.~\ref{structure} shows schematics of V-V pairing of VO$_{2}$ in different phases. Another intermediate monoclinic phase (M2) with only half of the V$^{4+}$ dimerized and the other half canting was first reported in Cr doped VO$_{2}$\cite{Marezio72,Pouget74}. Later, the M2 phase was also found to be stable under certain conditions, for example, by applying very small uniaxial stresses to pure VO$_{2}$\cite{Pouget75} or other transition metal substitutions involving lower oxidation states\cite{Strelcov12}. The uniaxial stress measurements were exceptionally significant for two very different reasons. On the fundamental side they demonstrated that in pure VO$_{2}$, at 340 K, there is a near degeneracy of the R, M1 and M2 phases. This observation has been used to argue that argue that VO$_{2}$ is as clear example of a Mott-Hubbard insulator and can also be used to argue that VO$_{2}$ is an example of a boot-strapped spin-Peierls transition. On the applied/operational side, the profound strain sensitivity of VO$_{2}$ requires strain free samples for measurements of intrinsic properties and offers the possibility of using strain, e.g. in thin films via epitaxial mismatch, to tune/modify the system.

VO$_{2}$ doping with Ti has been demonstrated to be one of the ways to stabilize the M2 phase in between the R and M1 phases at remarkably low Ti doping levels. However, so far, samples have been primarily studied in thin film and polycrystalline form\cite{Beteille97,Horlin76,Hiroi13}. In this paper, we present the details of how to grow pure and Ti-doped single crystals of VO$_{2}$ in as low strain of a state as possible. Given the profound sensitivity of VO$_{2}$ to strain, the availability of such samples is vital for providing intrinsic, bulk comparisons to the growing number of thin film studies of pure and doped VO$_{2}$. In addition, we demonstrate the effect of Ti-valence on doping level when using solution growth out of V$_{2}$O$_{5}$.

\begin{figure}[!ht]

\includegraphics[width=0.9\textwidth]{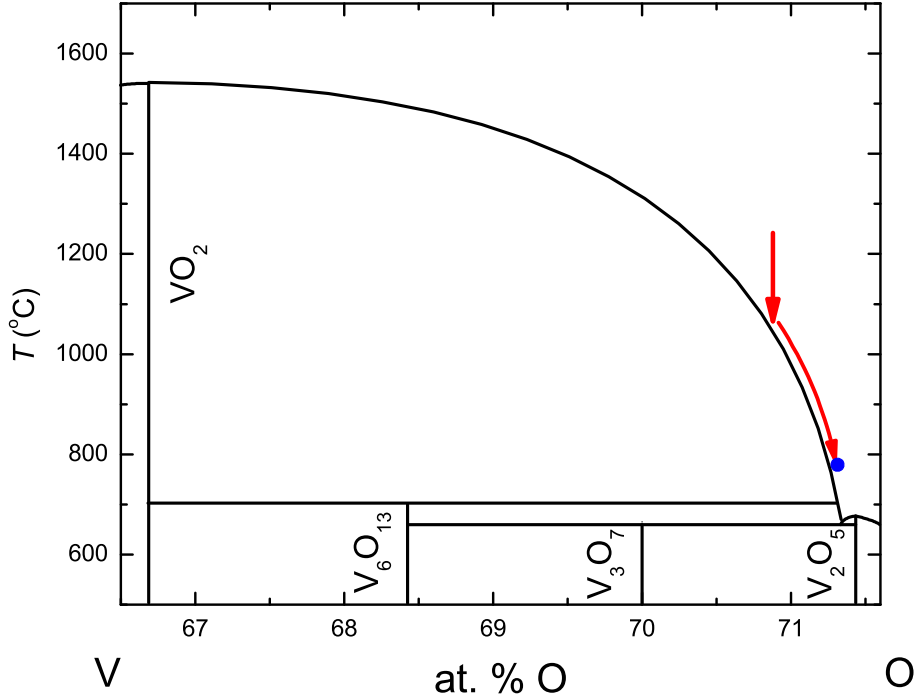}

\caption{(Color online) A schematic V-O binary phase diagram. The starting stoichiometry and the part of the phase diagram that is used to grow VO$_{2}$ is indicated by the red arrows. The blue point indicates the decanting point.}
\label{diagram}
\end{figure}

\begin{figure}[!ht]

\includegraphics[width=0.9\textwidth]{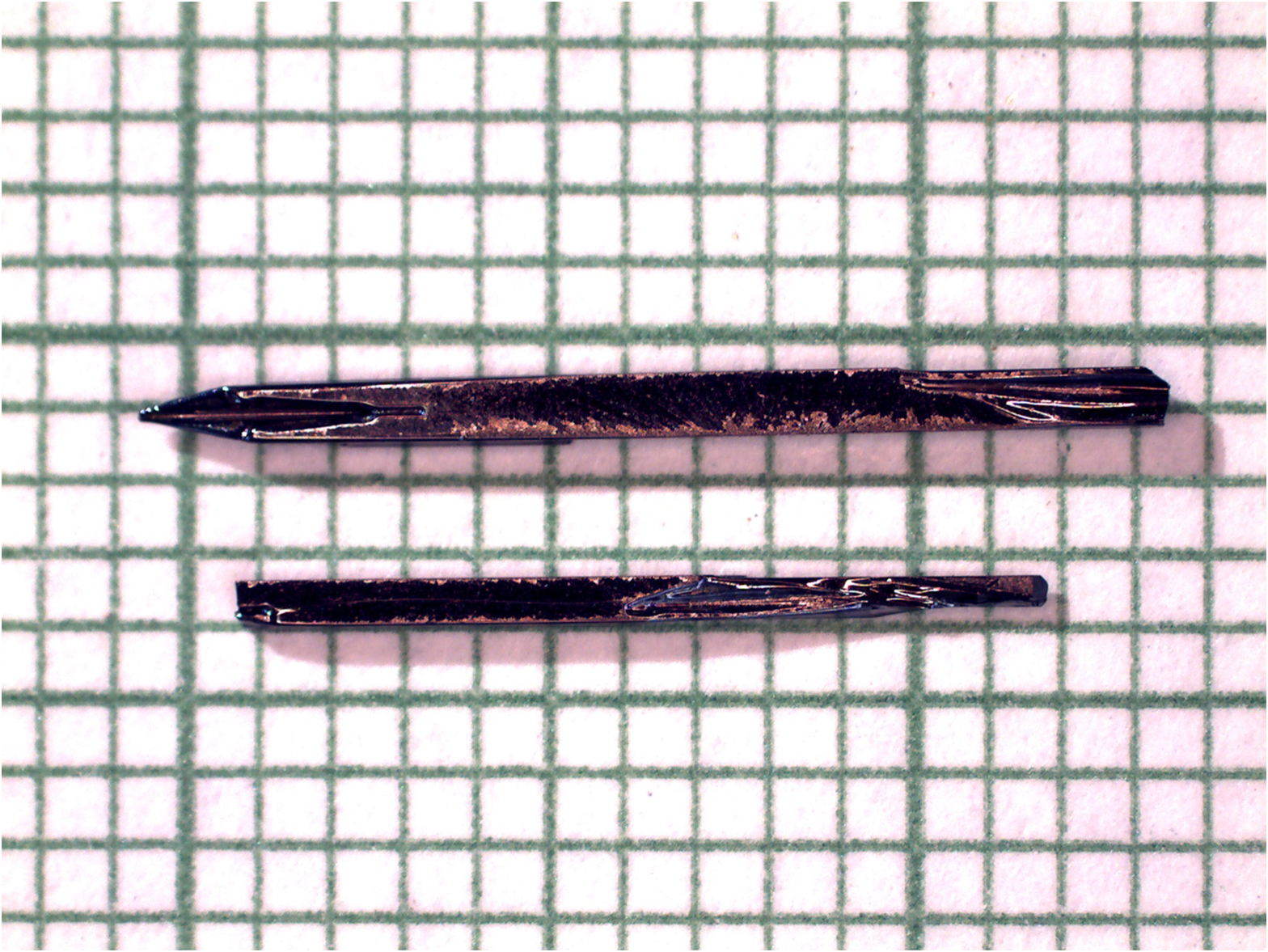}

\caption{(Color online) Typical single crystals of pure VO$_{2}$ on a millimeter grid paper. Note small amounts of solidified V$_{2}$O$_{5}$ flux along some edges.}
\label{xtal}
\end{figure}

Single crystals of V$_{1-x}$Ti$_{x}$O$_{2}$ were grown using a high-temperature solution growth technique\cite{Canfield92,Allen93}. Typical starting materials for a pure VO$_{2}$ growth were roughly 1 gram of VO$_{2}$ lump, which was obtained by reducing V$_{2}$O$_{5}$ in a N$_{2}$ atmosphere, and 8.1 grams of V$_{2}$O$_{5}$ powder. The sealed silica tube that holds the mixture of materials was heated up to 1050 $^{\circ}$C and slowly cooled over up to 100 hours to 775 $^{\circ}$C, at which temperature the remaining liquid was separated from the single crystals through a quartz wool plug via centrifugation\cite{Canfield92}. Fig.~\ref{diagram} presents a schematic V-O binary phase diagram with the part of the phase diagram that is used to grow VO$_{2}$ and decanting temperature marked by red arrows and a blue point respectively. TiO$_{2}$ powder was added into the VO$_{2}$, V$_{2}$O$_{5}$ mixture to obtain various V$_{1-x}$Ti$_{x}$O$_{2}$ samples. Typical single crystals are needle-like as shown in Fig.~\ref{xtal}. With increasing Ti doping, the crystals get thinner.

Powder X-ray diffraction was measured using a Rigaku Miniflex II desktop X-ray diffractometer (Cu K$_{\alpha}$ radiation). Magnetization was measured using a Quantum Design (QD) Magnetic Property Measurement System (MPMS) superconducting quantum interference device (SQUID) magnetometer. Elemental analysis was performed by Wave Dispersive Spectroscopy (WDS) in the electron probe microanalyser of a JEOL JXA-8200 electron microprobe.

\begin{figure}[!ht]

\includegraphics[width=0.9\textwidth]{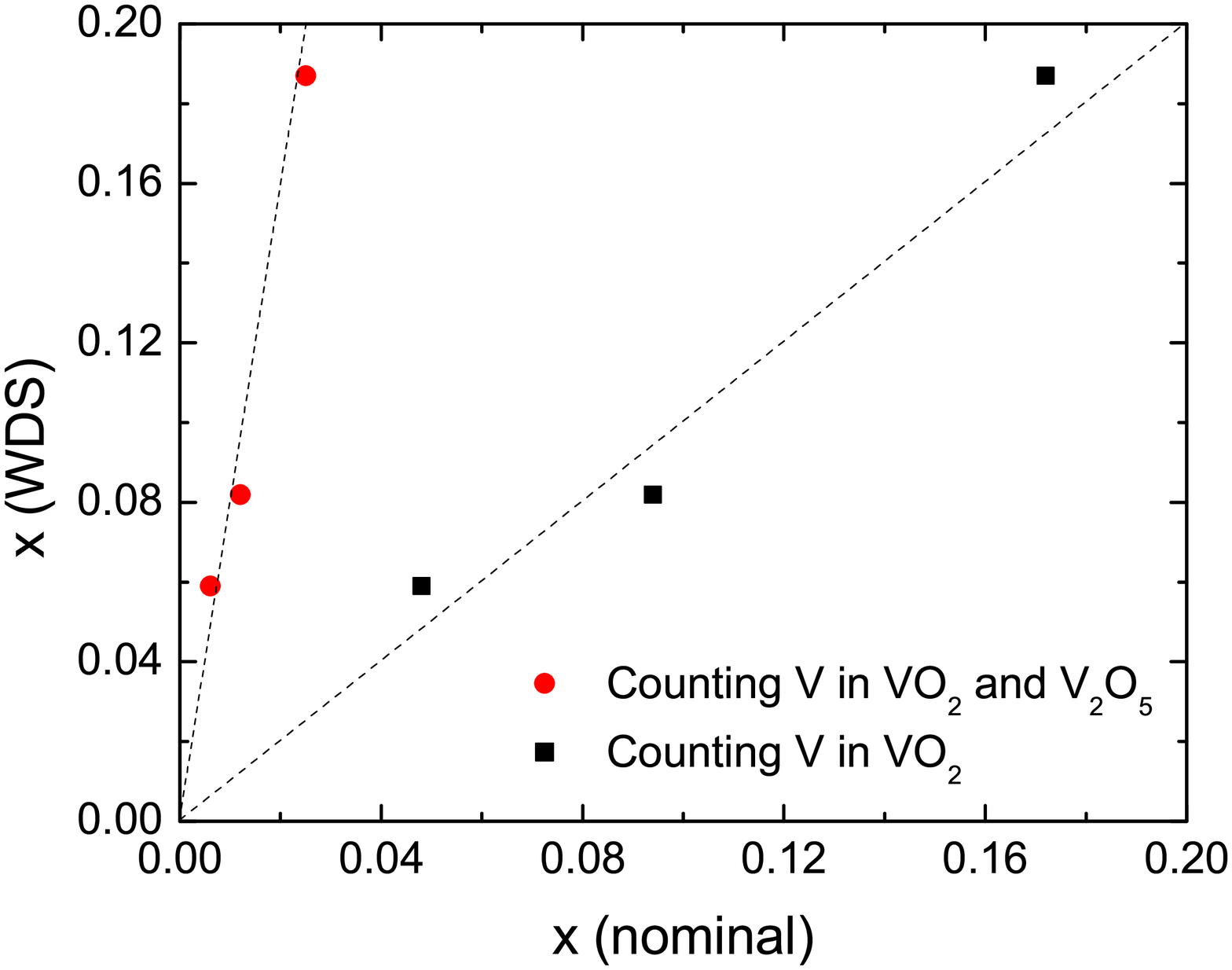}

\caption{(Color online) The WDS determined Ti concentration, x (in V$_{1-x}$Ti$_{x}$O$_{2}$), versus x-nominal. The x-nominal values represented by the black squares were determined by only considering the amount of V$^{4+}$ in the starting melt. Red circles take into account of all the V in the starting melt.}
\label{wds}
\end{figure}

The Ti-concentration in the V$_{1-x}$Ti$_{x}$O$_{2}$ single crystals was determined via WDS analysis. Fig.~\ref{wds} plots the x-WDS value versus the nominal value of Ti in the high-temperature melt. It is important to point out that two methods of evaluating the nominal Ti doping level are shown in Fig.~\ref{wds}. The red circles plot the x-WDS versus the x-nominal value determined by comparing the Ti level to the total amount of V in the melt. In an intermetallic growth this would be considered to be the standard manner of determining the x-nominal value. As can be seen there is a roughly linear dependence of x-WDS versus x-nominal, but the slope is close to eight. The black squares plot x-WDS versus x-nominal value determined by comparing the Ti level to the V$^{4+}$ level in the melt (i.e. comparing the Ti$^{4+}$ level from the TiO$_{2}$ to the V$^{4+}$ level from the VO$_{2}$). As can be seen in this case the data fall very close to a line with a slope of unity. This result makes sense considering that Ti cannot have a higher oxidation level and is essentially trapped in the Ti$^{4+}$ state by stoichiometry and the V$_{2}$O$_{5}$ melt. x$_{WDS}$ values are used throughout this paper to identify the samples.

\begin{figure}[!ht]

\includegraphics[width=0.9\textwidth]{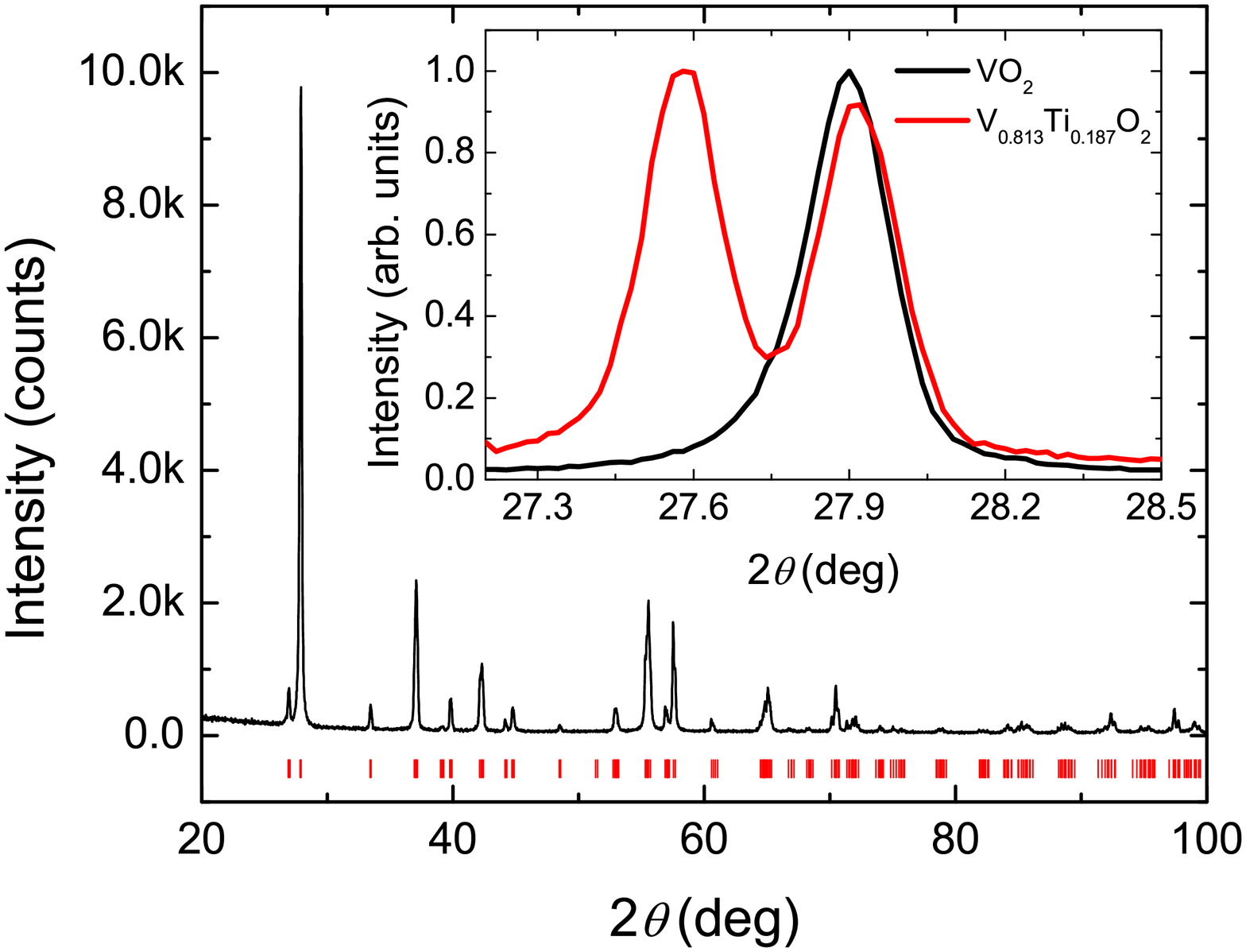}

\caption{(Color online) The powder diffraction pattern from pure VO$_{2}$ with theoretical peak positions indicated by ticks at the bottom. Inset shows the comparison between VO$_{2}$ (black) and V$_{0.813}$Ti$_{0.187}$O$_{2}$ (red) at 2$\theta$ $\sim$ 28 $^{\circ}$.}
\label{xray}
\end{figure}

Fig.~\ref{xray} shows the room-temperature powder X-ray diffraction data of the pure VO$_{2}$ over a 2$\theta$ range of 20-100$^{\circ}$. All peaks can be fitted to the M1 monoclinic structure of VO$_{2}$. Upon doping with Ti, the M2 phase\cite{Horlin76,Hiroi13} appears in between the R and M1 phases. The phase boundaries between R-M2 and M1-M2 split with increasing amounts of Ti substitution, as is shown in Fig.~\ref{summ} below. Above about 15$\%$ of Ti substitution, the M1-M2 phase boundary is pushed below room-temperature. In the inset of Fig.~\ref{xray}, we show a characteristic peak at around 2$\theta \sim$ 28$^{\circ}$ from both VO$_{2}$ and V$_{0.813}$Ti$_{0.187}$O$_{2}$. The clear splitting of the diffraction peak in the doped sample is consistent with such a change in crystal structure.

\begin{figure}[!ht]

\includegraphics[width=0.9\textwidth]{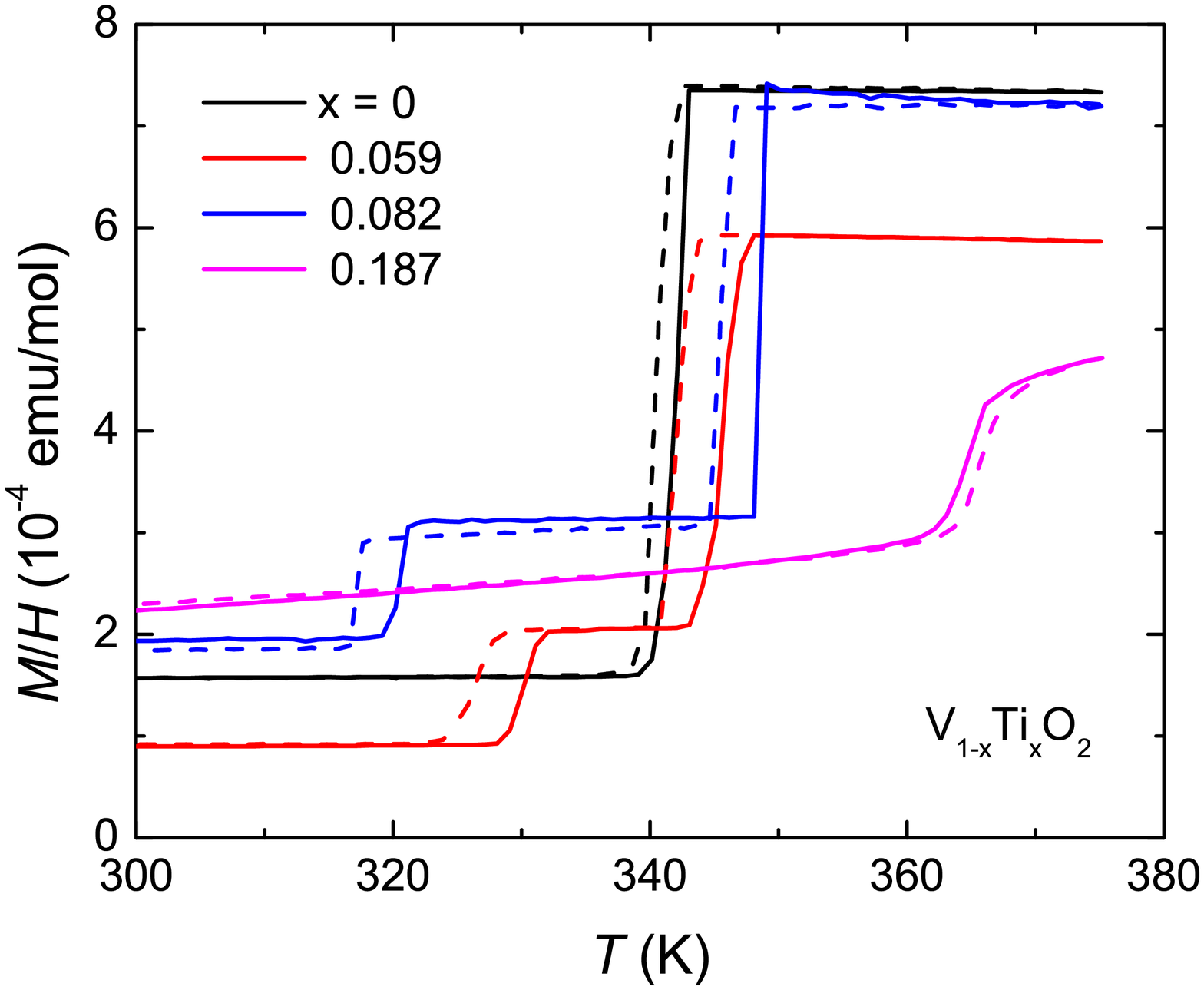}

\caption{(Color online) The temperature-dependent dc magnetic susceptibility of V$_{1-x}$Ti$_{x}$O$_{2}$ measure at 10 kOe. The magnetization values shown here contain a small amount of diamagnetic signal from the sample holder (see text). Solid and dotted lines represent data obtained on warming and cooling respectively. Note: for x = 0.187, data were collected down to $T$ = 200 K and no signature of the M1-M2 transition was found.}
\label{Ti}
\end{figure}

Temperature-dependent dc magnetic susceptibilities measured both on cooling and warming in 10 kOe are presented in Fig.~\ref{Ti}. For most of the measurements, a transparent plastic capsule was used to hold a collection of crystalline rods (see Fig.~\ref{xtal}) in order to acquire a large enough signal. Therefore, apart from pure VO$_{2}$, the data shown in Fig.~\ref{Ti} also contains a small diamagnetic background signal from the sample holder. It should be pointed out that although Fig.~\ref{Ti} plots data between 300 K and 375 K for clarity, for x = 0.187, data were collected down to $T$ = 200 K and no signature of a lower-temperature transition was found.

\begin{figure}[!h]

\includegraphics[width=0.9\textwidth]{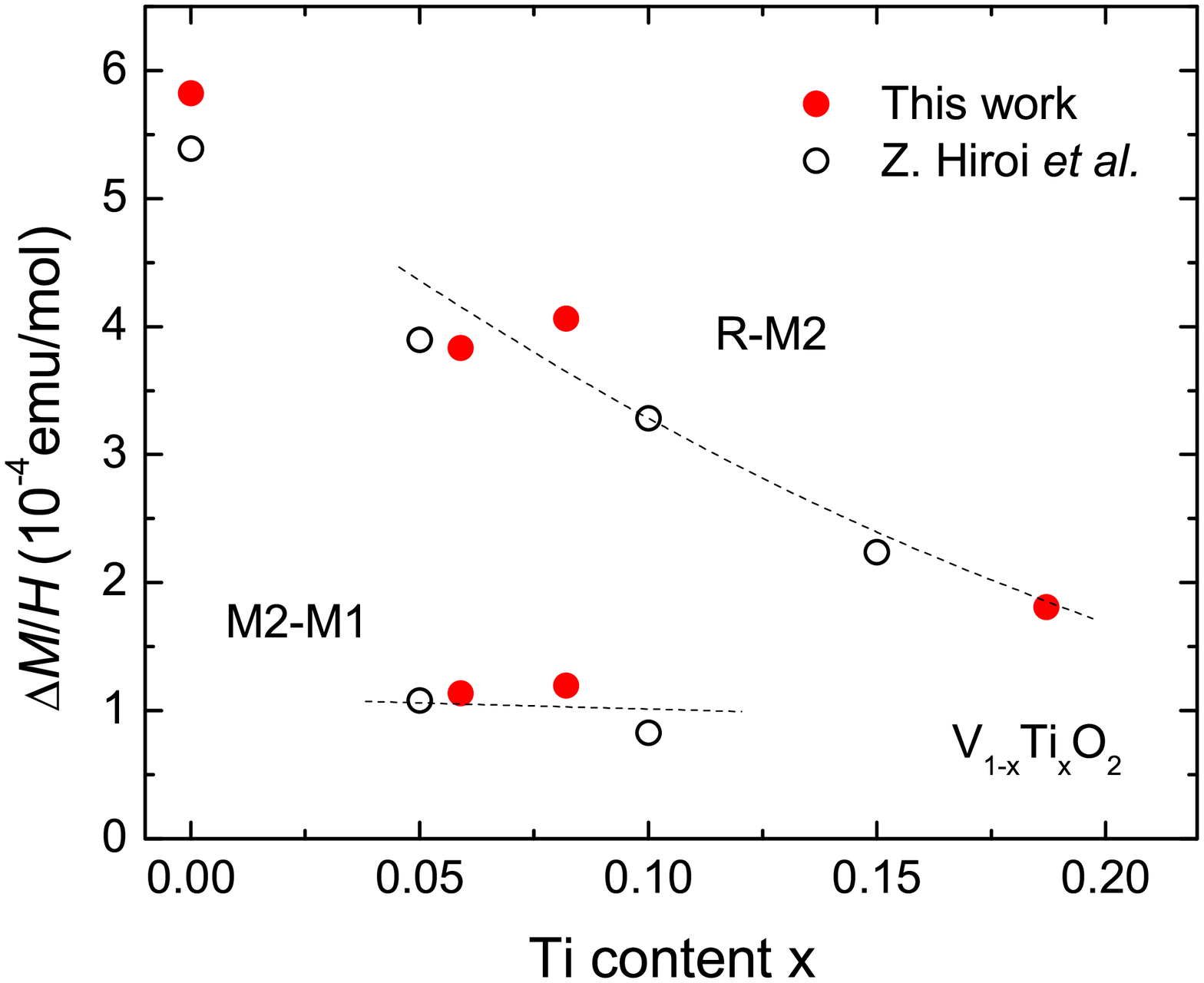}

\caption{(Color online) The change in magnetic susceptibility at the R-M1 transition for the pure VO$_{2}$ and R-M2, M2-M1 transitions for Ti doped VO$_{2}$ samples are plotted in red dots. Black circles represent the experimental data from polycrystalline samples\cite{Hiroi13}. Dashed lines are guides for the eye.}
\label{deltaM}
\end{figure}

A sharp first-order transition is clearly observed in VO$_{2}$ at $\sim$ 340 K, which corresponds to the metal-insulator, structural R-M1, phase transition. In comparison, with Ti doping such as x = 0.059 and 0.082, the single, first-order transition splits into two, sharp, well defined, first-order transitions. In between these two, first-order transitions, the M2 phase is stabilized\cite{Horlin76,Hiroi13}. Taking the peak positions of the derivatives of the temperature-dependent magnetization as transition temperature values, the evolution of the transition temperatures can be plotted as a function of Ti concentration. In Fig.~\ref{summ}, the transition temperatures obtained in this study are plotted together with recent results from a study of polycrystalline samples\cite{Hiroi13}. Since the sample holder's signal is essentially temperature-independent over this temperature range, we can also look at the magnetic susceptibility change associated with each phase transition. Fig.~\ref{deltaM} shows the size of magnetic susceptibility jump at each transition plotted as a function of Ti substitution level.

\begin{figure}[!h]

\includegraphics[width=0.9\textwidth]{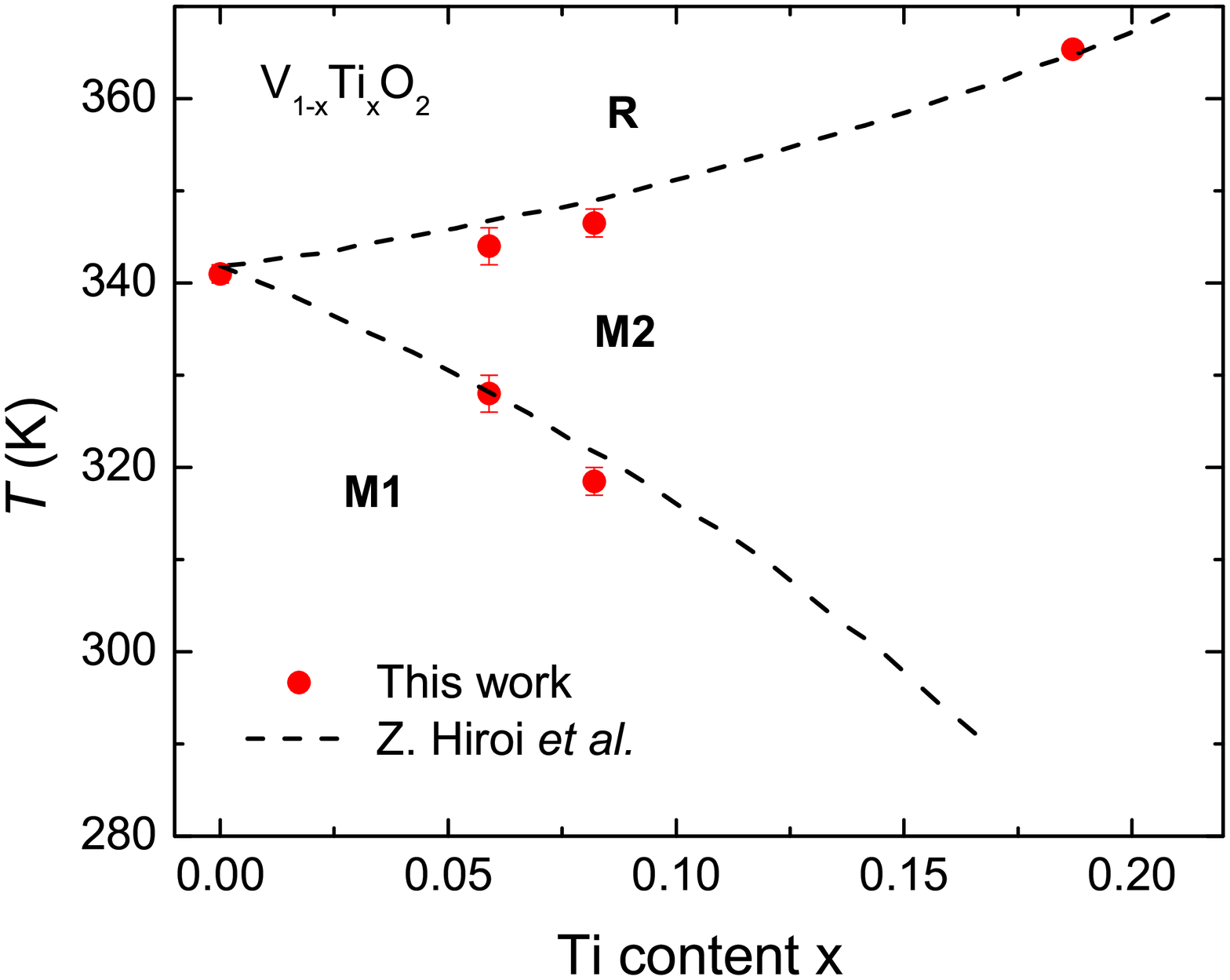}

\caption{(Color online) The magnetic transition temperature as a function of Ti doping ratio in VO$_{2}$. Error bars cover the range of transition temperatures obtained both on cooling and warming. Black lines show the trend from the polycrystalline study\cite{Hiroi13}. Note: no signature of the M1-M2 transition was found down to 200 K for x = 0.187.}
\label{summ}
\end{figure}

With increasing Ti substitution, the R-M2 transition temperature moves higher while the M1-M2 phase transition temperature moves lower (Fig.~\ref{summ}). In between, the M2 phase is stabilized over a larger and larger temperature range. In the M2 phase, half of the V$^{4+}$ ions are dimerized and thus show an intermediate level of magnetic susceptibility [see Fig.~\ref{structure}(c)]. The loss of magnetization at the R-M2 phase transition decreases at a rate of roughly 0.2$\times$10$^{-4}$ emu/mol per 1$\%$ Ti doping. On the other hand, the change in magnetization at the M1-M2 transition remain roughly unchanged with respect to the amount of Ti substitution. The total loss of magnetization from the R phase to M1 decreases with increasing amount of Ti. This can be roughly understood as a consequence of replacing magnetic V$^{4+}$ with non-magnetic Ti$^{4+}$. Ti substitution results in a decrease in magnetization in the paramagnetic R phase by reducing the amount of V$^{4+}$, and an increase of magnetization in the non-magnetic M1 phase by increasing the amount of un-paired V$^{4+}$ ions. The un-paired V$^{4+}$ also give rise to a clear Curie tail at low temperatures\cite{Hiroi13}. It worth pointing out, however, by looking at Fig.~\ref{deltaM}, that the magnetization loss at the R-M2 phase transition is larger than that at the M1-M2 transition, conflicting with the simplified picture of the M2 phase being associated with a pairing of half of the V$^{4+}$ from the higher temperature, paramagnetic VO$_{2}$. This might indicate that the V$^{4+}$ ions that sit in the neighborhood of Ti$^{4+}$ ion tend not to dimerize but rather form the zigzag type structure. Assuming a homogeneous Ti substitution, this disruption of the dimer formation will bring increasing disorder and defects to the M1 and M2 structures. At higher Ti substitution levels, this disorder may be the cause of the broadening of the R-M2 transition as well as the absence of the M1-M2 transition for the x = 0.187 sample.

Both Fig.~\ref{summ} and Fig.~\ref{deltaM} show excellent agreement between our single crystal data with the polycrystalline data from Ref.\onlinecite{Hiroi13}. This agreement indicates that in the polycrystalline samples, unlike thin film samples, strain is not playing a significant role\cite{Yang11}.


In conclusion, we've used a high-temperature solution technique to grow large, low strain, single crystals of V$_{1-x}$Ti$_{x}$O$_{2}$ (0 $<$ x $<$ 0.187). For Ti-substitution this growth technique clearly segregates the transition metal ions by valence and the substitution level of the V$_{1-x}$Ti$_{x}$O$_{2}$ crystals is most clearly related to the ratio of Ti$^{4+}$ : V$^{4+}$ in the melt rather than to the total Ti : V ratio. Phase transition temperatures were determined by temperature-dependent dc magnetization measurements. The R-M2 (M1-M2) phase transition temperature increases (decreases) with Ti doping. The size of the magnetic susceptibility change at each transition was studied. The fact that the magnetic susceptibility change at R-M2 and M1-M2 are not equivalent suggests that there may be a preference for pairing V$^{4+}$ ions (rather than Ti$^{4+}$-V$^{4+}$ ions) in the M2 phase. A systematic temperature-dependent X-ray diffraction study will be needed to provide more details about the Ti doping effect on the structure and stability of VO$_{2}$ in these phases.

The authors would like to thank W. E. Straszheim, S. M. Saunders for experimental assistances and R. J. Cava for insightful discussions. This work was supported by the U.S. Department of Energy (DOE), Office of Science, Basic Energy Sciences, Materials Science and Engineering Division. The research was performed at the Ames Laboratory, which is operated for the U.S. DOE by Iowa State University under contract NO. DE-AC02-07CH11358.

\bibliographystyle{apsrev4-1}
%

\end{document}